\documentclass[aps, pra, a4paper, 10pt, showpacs, twocolumn]{revtex4-1}
\usepackage{amsmath, amssymb, amsthm, bm, bbm, textcomp}
\usepackage[T1]{fontenc}
\usepackage[latin9]{inputenc}
\usepackage{graphicx,graphics}
\usepackage{geometry}
\usepackage{color}
\newcommand{\ket}[1]{\left|{#1}\right\rangle}
\newcommand{\bra}[1]{\left\langle{#1}\right|}

\newcommand{\be}{\begin{equation}}
\newcommand{\ee}{\end{equation}}
\newcommand{\eea}{\end{eqnarray}}
\newcommand{\bea}{\begin{eqnarray}}

\begin{document}

\title{Measurement-based implementation makes entanglement purification based on hashing practical}

\author{M. Zwerger$^{1,2}$, H. J. Briegel$^{1,2}$, and W. D\"ur$^1$}
\affiliation{
$^1$ Institut f\"ur Theoretische Physik, Universit\"at Innsbruck, Technikerstra{\ss}e 25, A-6020 Innsbruck,  Austria\\
$^2$ Institut f\"ur Quantenoptik und Quanteninformation der \"Osterreichischen Akademie der Wissenschaften, A-6020 Innsbruck, Austria}

\begin{abstract}
We investigate entanglement purification protocols based on hashing, where a large number of noisy entangled pairs is jointly processed to obtain a reduced number of perfect, noiseless copies. While hashing and breeding protocols are the only purification protocols that asymptotically obtain a nonzero yield, they are not applicable in a realistic scenario if local gates and measurements are imperfect. We show that such problems can be overcome by a compact measurement-based implementation, yielding entanglement purification schemes with nonzero yield that are applicable also in noisy scenarios, with tolerable noise per particle of several percent. We also generalize these findings to multiparty purification protocols for arbitrary graph states.
\end{abstract}

\date{\today}

\maketitle

\section{Introduction}
Entanglement purification is an important primitive in quantum information processing \cite{Be96,Be96b,Du07}. This follows from the fact that entanglement is a key resource in quantum communication and quantum computation, and the generation and distribution of high-fidelity entangled states is of central importance. Entanglement purification provides a possibility to achieve this aim, even in the presence of noise and imperfections in the local operations and the apparatus used for the purification. In its initial form, entanglement purification protocols were introduced to reduce or circumvent channel noise. By sending parts of a (locally generated) maximally entangled pair through a noisy channel, one ends up with distributed noisy entangled pairs shared between two parties. From many such noisy copies, fewer copies with increased fidelity or even unit fidelity can eventually be generated by means of local operations, i.e. operations that act on several copies but are performed locally by individual parties. The resulting entangled states can then be used e.g. for quantum communication via teleportation \cite{Be93}, for quantum key distribution \cite{Ek91}, or for performing remote operations \cite{Be01,Be05}.

Various kinds of entanglement purification protocols have been proposed. They differ in the number of pairs they act on, and they may be deterministic or probabilistic (see e.g. \cite{Du07}). From a practical perspective, recurrence protocols \cite{Be96,De96} that operate on two copies are most important. In such a protocol, one pair is measured to reveal non-local information about the other pair. Only for certain measurement outcomes, the fidelity of the remaining pair is increased. A recursive application brings the resulting states closer and closer to maximally entangled states, however unit fidelity is only approached asymptotically. Due to the fact that in each step at least one of the two pairs is discarded, the yield of the procedure (i.e. fraction of perfect pairs that can be obtained from many noisy pairs) tends to zero. In contrast hashing (and also breeding) protocols \cite{Be96,Be96b} operate on an infinite ensemble of identical copies of noisy entangled pairs in the first place, and only a finite fraction of the pairs is measured to reveal information on the remaining ensemble. For sufficiently high initial fidelity, the remaining pairs are in a maximally entangled state, and therefore such a protocol has nonzero yield.

However, hashing protocols have a very serious drawback: they are not applicable in a realistic scenario. If local operations and measurements are noisy as well -as is the case in any practical implementation- it is straightforward to see that the protocols fail - even if the amount of noise is arbitrarily small \cite{Du07}. The reason for this is that these protocols operate globally - that is, in order to obtain information about a large fraction of the ensemble (e.g. its parity), operations on this large fraction need to be performed and the information to be encoded into some of the pairs. The measurement on one pair of the ensemble reveals only one bit of information. However, each of the noisy operations increases the entropy of the ensemble, and as there are $O(N)$ two-qubit operations required \cite{Be96b}, the increase in entropy due to noisy operations exceeds the information gain (entropy reduction) due to the measurement even for tiny imperfections, thereby jeopardizing the whole protocol.

Here we show that these problems can be circumvented if hashing protocols are implemented in a measurement-based way \cite{Zw12,Zw13}. That is, rather than performing sequences of gates on the noisy ensemble, certain entangled resource states are prepared locally by each of the parties, and coupled via Bell measurements to the particles of the ensemble to be purified. Since all involved operations for hashing protocols are of so-called Clifford type, the size of the resource state is $N+M \leq 2N$ for any such protocol that operates on $N$ input pairs and produces $M$ output pairs, even though the required number of operations in the circuit model is $O(N^2)$. Sources for noise in this case are an imperfect preparation of resource states, and imperfect Bell measurements. We find that noise on the resource state of up to 7\% {\em per particle} is tolerable. We remark that only the reduced size of the resource states makes the protocol practical, c.f. for similar observations for recurrence protocols \cite{Zw13}. Clearly, the achievable fidelity is smaller than unity, however the yield is still nonzero. Notice that a (concatenated) measurement-based implementation of an entanglement purification scheme based on a recurrence protocol has the problem that the success probability --and hence the yield of the protocol-- drops exponentially with the system size (which follows from the fact that all purification steps have to be successful simultaneously). This can be avoided by using the hashing scheme we propose here.

The paper is organized as follows. In Sec. \ref{hashing} we provide background information on hashing and recurrence protocols, and their measurement-based implementation. In Sec. \ref{results} we provide error thresholds for the measurement-based protocols and extend our results to the purification of multipartite graph states. Finally we summarize and conclude in Sec. \ref{discussion}.

\section{Background}
\label{hashing}

The hashing protocol \cite{Be96,Be96b} is a particular entanglement purification protocol operating on an infinite ensemble of noisy entangled pairs. The goal is to obtain $M$ Bell pairs with asymptotically unit fidelity from a larger number $N$ of impure Bell pairs drawn from a Bell diagonal ensemble. The protocol is based on parity measurements of subsets of the ensemble, which provide information about the remaining/unmeasured pairs. Thereby, $N-M$ such measurement rounds are conducted. The size of the subsets in each round is of the order of the system size, $O(N)$, and each round involves $O(N)$ controlled phase gates (CZ $=diag(1,1,1,-1)$ in computational basis) and some single qubit operations. These operations allow one to transfer the information about the parity of the subset to a target pair, which is ultimately measured to reveal it. It turns out that the number $M$ of distillable perfect Bell pairs approaches $M \approx N(1 - S(W))$ in the asymptotic limit, where $S(W)$ denotes the von Neumann entropy of the impure Bell pairs in the original ensemble. This is a key difference between the hashing and recurrence protocols, which have vanishing yield $D=\tfrac{M}{N}$ in the asymptotic limit. This follows from the fact that in each purification step one of the pairs is measured to obtain information (a similar argument applies for $n \rightarrow m$ protocols, where in each purification step $n$ input pairs are mapped to $m$ output pairs, and the remaining ones are measured). Improvements of the yield and generalizations of the original hashing and breeding protocol have been found in \cite{Vo05,Ho06,Vo03,Ho04,Gh05,De05}.

Hashing protocols have also been introduced for GHZ states \cite{Ma02}, two-colorable graph states \cite{As05b} and general graph states \cite{Kr06}.

One drawback of hashing protocols compared to recurrence protocols is that they can not tolerate noise in the operations, i.e. imperfections in the gates and measurements themselves. Whereas it is known that recurrence protocols can tolerate several percent of noise, the hashing protocols collapse for arbitrarily small amount of noise in the asymptotic limit (see e.g. \cite{Du07} and the discussion above).

\section{Results}
\label{results}

In this section we show how one can construct the resource states for measurement-based entanglement purification and how noise is modeled. Furthermore we derive the error threshold for the resource states which can be used to purify Bell pairs as well as for 1D and 2D cluster states \cite{Ra01b}.

\subsection{Resource states}

The purification protocols have $n$ input and $m$ output qubits (for each party) and use only Clifford gates and Pauli measurements. Consequently, one can implement them with resource states containing $n+m$ qubits \cite{Ra03}. These states can be constructed in different ways. One possibility is to start with a 2D cluster state, which is a universal resource state for measurement-based quantum computation \cite{Ra01,Br09}, and the measurement pattern for the protocol, which will contain only Pauli measurements. The state which results after applying all these measurements can be calculated using the rules for Pauli measurements on graph states \cite{He04,He06}.

A different approach is to make use of the Jamiolkowski isomorphism \cite{Ja72}, which relates a completely positive map with a state. This state is obtained by applying the map to $n$ qubits, each of which is part of a Bell states $\ket{\phi^+} = 1/\sqrt{2} \left(\ket{00} + \ket{11}\right)$, assuming that the map acts on $n$ qubits.

In both cases the calculation can be carried out efficiently on a classical computer, which follows from the Gottesman-Knill theorem \cite{Go97,Aa04,An06}.

The protocol is then implemented by coupling the (unknown) input states which shall be purified to the resource states via Bell measurements. Depending on the measurement results one has to deal with byproduct operators, which can be commuted through the circuit encoded in the resource state. Notice that this only works if the circuit contains only Clifford gates and Pauli measurements. In addition, the measurement results reveal the information about the parities of the various subsets (this is similar to the measurement-based implementation of recurrence protocols which is discussed in more detail in \cite{Zw12}).

The influence of imperfections is modeled by local depolarizing noise (LDN). LDN can be seen as a worst case scenario for local noise, because any local noise can be brought to this form \cite{Du05}. Given a pure $n$ qubit state $\rho = \ket{\psi}\bra{\psi}$, the noisy density matrix reads
\be
\rho_{noisy} = \prod_{j=1}^n {\cal{D}}_j(p) \rho,
\ee
with ${\cal{D}}_j(p) \rho = p\rho + \tfrac{1-p}{4} (\rho + X_j\rho X_j + Y_j\rho Y_j + Z_j\rho Z_j)$. Here, $X$, $Y$ and $Z$ are the Pauli operators and the subscript indicates on which subsystem they act. The Bell measurements used to read-in the input states can be assumed to be perfect since imperfections can be included in the noise parameters of the graph states, leading to new, lower values of $p$. Notice, that the fidelity of the resulting states drops exponentially with the system size within this error model (in leading order in $(1-p)$), $F \approx \left(\tfrac{3p + 1}{4} \right) ^n$.

\subsection{Error thresholds}

In \cite{Zw13} it was shown that one can exchange the location of LDN when followed by a Bell measurement. To be precise,
\be
{\cal{P}}_B {\cal{D}}_1 \rho =  {\cal{P}}_B {\cal{D}}_2 \rho
\ee
with ${\cal{P}}_B \rho = P_B \rho P_B ^{\dagger}$ where $P_B$ is a projector on a Bell state of particle 1 and 2. This allows one to derive the error thresholds in a fairly simple way. Assume that the Bell pairs, or more generally the graph states which one would like to purify, are affected by LDN with parameter $q$, and the resource states (that are used to implement the purification) by LDN with parameter $p$. Then one can use the property described above to effectively move the noise from the resource state to the incoming Bell pairs, which will then be described by LDN with parameter $pq$, as follows from ${\cal{D}}(p){\cal{D}}(q) = {\cal{D}}(pq)$. The noise acting on the output qubits of the resource state can be applied in the last step. The advantage of this decomposition is that, after moving the noise, one deals with a perfect purification for which analytical results are available. In the asymptotic limit the protocols simply output a perfect Bell pair. The conditions for purification to work are then the following ones. First, the LDN parameter of the Bell pairs, after the noise from the resource state is moved, has to be larger than the minimal required value $q_{min}$ such that purification is possible, i.e.
\be
pq > q_{min}.
\ee
Second, the LDN parameter $p$ of the final Bell pairs has to be larger than the parameter $q$ of the incoming pairs,
\be
p > q.
\ee
The threshold value $p_{min}$, such that purification is possible for $p > p_{min}$ is then given by $p_{min} = \sqrt{q_{min}}$.

The derivation of the error thresholds for a measurement-based entanglement purification protocol in the asymptotic limit thus reduces to determine the minimal value $q_{min}$ such that a Bell pair or graph state can be purified, using this protocol. This will be done in the following subsection for the cases of Bell pairs as well as 1D and 2D cluster states.

\subsubsection{Bell pairs}

It has been shown in \cite{Be96b} that Bell pairs in Werner form with a fidelity exceeding $F_{min} \approx 0.8107$ can be purified with a hashing protocol. This translates to a value $q_{min} \approx  0.8672$. Consequently the resource state capable of implementing the hashing protocol can tolerate $1 - p_{min} \approx 6.9\%$ noise per particle. It is remarkable that several percent of noise per particle are tolerable in such a measurement-based implementation, while any tiny amount of noise in a gate-based implementation renders the protocol impractical. The reason for this is that the measurement-based implementation allows for a significant reduction of the size of the resource state in the following sense. The gate-based implementation of the hashing protocol involves $O(N^2)$ gates, which would lead to a resource state size of $O(N^2)$. However, since all gates are of Clifford type, a reduction of the size to $N+M \leq 2N$ is possible. 

This threshold value is lower than values observed in \cite{Zw13} for other measurement-based entanglement purification protocols. This  is a direct consequence of the required high value of $F_{min}$ for the Hashing protocol to work, while other protocols (such as recurrence protocols) are known to only require a minimal fidelity of $F_{min,R}=0.5$. The required value of $F_{min} \approx 0.8107$ can be circumvented if one first uses a recurrence protocol, for which $F_{min,R}=0.5$, to increase the fidelity of the ensemble and then switches to the hashing protocol. However, this does not affect the error threshold.

Furthermore, the usage of any type of recurrence or $n \to m$ purification protocol has another drawback. Although the error threshold is significantly higher, the overall success probability of such protocols is (exponentially) small. First, when combining several rounds of such protocols in a single step, all steps need to be successful simultaneously, yielding a success probability that decreases exponentially with the number of initial pairs. For a small number of such recurrence rounds, this is not a big problem. When attempting a combination of recurrence and hashing protocols in a measurement-based way, a second problem appears. Even if only one recurrence round is done before switching to hashing, all these initial purification steps need to be successful simultaneously over the whole ensemble, as the compact resource state does not allow to exclude certain pairs where the recurrence purification step was not successful from further processing. Hence the success probability drops again exponentially. This can be circumvented by implementing the recurrence purification step and the hashing purification step separately in a measurement-based way. This allows one to avoid the exponential drop in success probability, and to achieve at the same time that the initial fidelity of the pairs can be significantly lower than $F_{min}$, i.e. a larger amount of channel noise is tolerable. The threshold of the overall scheme is still given by the second step, i.e. the one of the hashing protocol and hence about $7\%$ of noise per particle are tolerable also in this approach.

\subsubsection{Cluster states}

Hashing protocols for two-colorable graph states were introduced in \cite{As05b}. A particularly important class of two-colorable graph states is the family of 2D cluster states, as it serves as a universal resource state for MBQC {\cite{Ra01,Ra01b}. The yield of these protocols is given by
\be
\label{yield}
D = 1-\operatorname{max}_{j \in V_A}\left[S(a_j^{(0)},a_j^{(1)})\right] - \operatorname{max}_{k \in V_B}\left[S(a_k^{(0)},a_k^{(1)})\right].
\ee
Here, $V_A$ and $V_B$ denote the two vertex subsets, corresponding to the two different colors. The entropy $S(a_j^{(0)},a_j^{(1)})$ is defined as
\be
S(a_j^{(0)},a_j^{(1)}) = -a_j^{(0)} \operatorname{log}_2 a_j^{(0)} - a_j^{(1)} \operatorname{log}_2 a_j^{(1)},
\ee
with $a_j^{(\mu_j)} = \sum_{\mu_k \neq \mu_j} \lambda_{\mu_1 \mu_2 \ldots \mu_j \ldots \mu_N}$, where the $\lambda_{\ldots}$ are the expansion coefficients in the graph state basis.

The $a_j^{(\mu_j)}$ depend on the LDN parameter $q$ and so does the yield $D$. One needs to determine the value $q_{min}$ such that the yield $D$ is non-vanishing for $q > q_{min}$. The equation (\ref{yield}) for the yield can be simplified if one considers infinitely large cluster states, or cluster states with periodic boundary conditions. Then the system is invariant under translations and the $S(a_j^{(0)},a_j^{(1)})$ are all equivalent. It is straightforward but lengthy to write down the analytic expressions for the $a_j^{(\mu_j)}$, the results can be found in the Appendix.

For the 1D (2D) cluster state one obtains $q_{min} \approx 0.9204$ $(0.9515)$, which translates to tolerable noise of $1 - p_{min} \approx 4.1\%$ $(2.5\%)$ for the corresponding resource states.

\section{Discussion}
\label{discussion}

In this work we have shown that a (optimized) measurement-based implementation of hashing protocols for entanglement purification works also in the presence of noise. This is a key difference compared to a gate based implementation, where no noise in the operations can be tolerated. The threshold value of $6.9\%$ noise for the resource state for the purification of Bell pairs is considerably lower than the value which we found for a concatenated implementation of recurrence protocols ($24.0\%$) \cite{Zw13}. However, it is worth to note that this value is already twice as high as the threshold value of a stepwise implementation of a recurrence protocol \cite{Zw12}. The main advantage of hashing protocols is that the yield, i.e. the ratio of the number of pairs with maximal reachable fidelity to the number of input pairs, is nonzero, whereas it vanishes for recurrence protocols. Similar results are found for multipartite entanglement purification.

The ability to perform entanglement purification in the presence of noise together with a nonzero yield makes the (optimized) measurement-based approach particularly appealing for practical applications.

\section{Acknowledgements}
This work was supported by the Austrian Science Fund (FWF): P24273- N16, SFB F40-FoQus F4012-N16.

\section{Appendix}

Here we determine the $a_j^{(\mu_j)}$ as a function of $q$. First we observe that $a_j^{(0)} + a_j^{(1)} =1$. The value of $a_j^{(1)}$ is simply the probability that index $\mu_j$ in $\lambda_{\mu_1 \mu_2 \ldots \mu_j \ldots \mu_N}$ takes the value one. This can happen due to noise acting on the graph state, for which originally (prior to the application of LDN with parameter $q$) only $\lambda_{00 \ldots 0}$ differs from zero. One needs to collect all possibilities and their probabilities which change the coefficients $\mu_j$ from zero to one.

An example in the case of the 2D cluster state is that a $Z$ error acts on vertex $j$ and no errors act on the neighborhood $N(j)$, the probability of which is given by
\be
P_{example}  =  \frac{1-\tilde{p}}{3} \tilde{p}^4,
\ee
with $\tilde{p}=\tfrac{3q+1}{4}$.

The expression for $a_j^{(1)}$ is given by
\begin{widetext}
\be
%\begin{split}
a_j^{(1)}  = 2\left(\frac{1 - \tilde{p}}{3} \right) \left[ \tilde{p}^2 + 2\tilde{p}\left(\frac{1-\tilde{p}}{3} \right) + \left(\frac{1 - \tilde{p}}{3} \right)^2 \right] + \left( \tilde{p} + \frac{1 - \tilde{p}}{3} \right) \left[ 4 \left(\frac{1 - \tilde{p}}{3} \right) \left(\tilde{p} + \frac{1 - \tilde{p}}{3} \right) \right]
%\end{split}
\ee
for the 1D cluster state and
\be
\begin{split}
a_j^{(1)} & =  2 \left(\frac{1 - \tilde{p}}{3} \right) \left[ \tilde{p}^4 + 4\tilde{p}^3 \left(\frac{1-\tilde{p}}{3} \right) +  4\tilde{p} \left(\frac{1-\tilde{p}}{3} \right)^3 + 6\tilde{p}^2 \left(\frac{1-\tilde{p}}{3} \right)^2 + \left(\frac{1-\tilde{p}}{3}\right)^4 + \right. \\
& \left. +24\left(\frac{1-\tilde{p}}{3}\right)^2 \left(2\left(\frac{1-\tilde{p}}{3}\right) \tilde{p} +\tilde{p}^2 + \left(\frac{1-\tilde{p}}{3}\right)^2 \right) + 16 \left(\frac{1-\tilde{p}}{3} \right)^4 \right] +\\
 & +\left( \tilde{p} + \frac{1-\tilde{p}}{3} \right) \left[ 8 \left(\frac{1 - \tilde{p}}{3} \right) \left(3\left(\frac{1-\tilde{p}}{3}\right) \tilde{p}^2 + 3 \left(\frac{1-\tilde{p}}{3} \right)^2\tilde{p} +\tilde{p}^3 + \left(\frac{1-\tilde{p}}{3} \right)^3 \right) + 32 \left(\tilde{p} + \frac{1-\tilde{p}}{3} \right) \left(\frac{1-\tilde{p}}{3}\right)^3 \right]
% & + 2\left(\frac{1-\tilde{p}}{3} \right) \left[ 24\left(\frac{1-\tilde{p}}{3}\right)^2 \left(2\left(\frac{1-\tilde{p}}{3}\right) \tilde{p} +\tilde{p}^2 + \left(\frac{1-\tilde{p}}{3}\right)^2 \right) + 16 \left(\frac{1-\tilde{p}}{3} \right)^4 \right]
 \end{split}
\ee
for the 2D cluster state.
\end{widetext}

\end{document}